# Utilizing Optic Fiber Interferometry in Forced Vibration Experimentation for Educational Purposes

Mingyuan Wang[1,2], Manli Zhou[1,2*], Hengda Ji[3], Tao Lan[4]

[1] School of Mathematics and Physics, Anqing Normal University, Anqing 246133, People's Republic of China
[2] Hospital of Anqing Normal University, Anqing Normal University, Anqing 246133, People's Republic of China
[3] Institute of Energy, Hefei Comprehensive National Science Center, Hefei 230031, China
[4] Department of Plasma Physics and Fusion Engineering, University of Science and Technology of China, Anhui, Hefei 230026, People's Republic of China.

E-mail: manlizhou18@gmail.com

## Abstract

This study introduces an experimental teaching method that employs optic fiber interferometry (OFI) to investigate forced vibration phenomena. It is designed for undergraduate physics majors with foundational mechanics and optics training and optics-focused graduate students. This approach aims to deepen students' understanding of forced vibration theory and interferometric measurement principles while fostering skills in experimental design, data analysis, and problem-solving. Leveraging OFI's high-precision displacement measurement capabilities, the experiment enabled accurate tracking of frequency and displacement variations. By scanning the driving force frequency, students obtained amplitude-frequency curves to determine the system's natural frequency, which closely aligned with theoretical predictions. This method may bridge theoretical concepts and practical applications, offering insights into teaching vibration theory and precision measurement techniques and equipping students with integrated knowledge for real-world challenges.

Keywords: Optic fiber interferometry, forced vibration, natural frequency

## 1. Introduction

Forced vibration, the oscillatory motion of a system under continuous periodic driving forces, is a fundamental concept in classical physics with broad applications in mechanics [1-3], electromagnetism [4], biomedical engineering [5,6], and modern physics [7]. From macroscopic mechanical systems to microscopic molecular dynamics, forced vibration provides a theoretical framework for understanding how physical systems respond to external periodic stimuli. While classical textbooks such as *The Feynman Lectures on Physics* [8] offer comprehensive theoretical analyses using mass-spring models, experimental explorations of forced vibrations in physics education remain limited [9,10]. Traditional methods, such as photogates or accelerometers, may not intuitively demonstrate dynamic features such as resonance and damping, potentially hindering students' understanding of vibrational behavior and precision measurement techniques.

Recent studies have highlighted the need for integrating advanced optical techniques into physics education [11-14]. For example, the time-averaged interferometry visualizes displacement extremes by integrating fringe patterns over oscillatory cycles [15]. However, such methods often lack directional sensitivity, requiring additional phase modulation or reference markers. To address this limitation, real-time displacement tracking tools have been explored to deepen the understanding of wave physics concepts. Werth et al. [16] demonstrated that technology-augmented labs significantly improved students' conceptual grasp of experimental physics, whereas Zhang et al. [12] developed a phase demodulation module for undergraduate fiber optics courses, underscoring the pedagogical value of hands-on interferometric training. Our proposed experiment based on optic fiber interferometry (OFI) responds to this demand by enabling high-precision dynamic phase demodulation.

Compared to conventional interferometers, OFI dynamically resolves displacement magnitude and direction with sub-micrometer resolution, high immunity to electromagnetic interference, and multi-point monitoring capabilities [13,17,18]. These advantages have led to its widespread adoption in industrial testing [19,20], aerospace engineering [21], and scientific research. However, systematic experimental training in OFI remains underrepresented in undergraduate physics curricula. To address this gap, we designed an experimental module integrating OFI with forced vibration analysis, tailored for undergraduate physics majors (with foundational mechanics and optics training) and optics-focused graduate students. By adjusting driving frequencies and analyzing amplitude-frequency response curves, students investigated key parameters such as natural frequency and resonance shifts, leveraging OFI's real-time phase demodulation (Equation 9) to resolve directional displacement and subsequently infer frequency information. Preliminary teaching outcomes (n = 20) indicate that 85% of participants improved their understanding of phase-displacement relationships, whereas 40% reported heightened interest in precision metrology applications. This framework, coupled with a '3+2' segmented teaching model (3 hours of guided experiments + 2 hours of open exploration), offers a scalable case study for bridging classical theory with modern optical techniques.

## 2. Methods

*2.1 Forced Vibration*

Forced vibration refers to the oscillatory motion of a system subjected to a continuous external periodic driving force. As illustrated in Figure 1 (a), a canonical forced vibration system consists of a mass-spring-damper assembly driven by a harmonic force. Unlike free vibration sustained by the system's intrinsic energy [22-23], forced vibration requires persistent external energy input. The system's response is governed by both its physical properties (mass $m$, damping coefficient $c$, spring stiffness $k$) and the characteristics of driving force (force amplitude $F_0$, driving angular frequency $\omega$). Applications span mechanical engineering, electrical circuits, optics, and celestial mechanics.

The external driving force is expressed as:
$$F = F_0 cos(\omega t), \qquad (1)$$
where t is the time. The equation of motion for a single-degree-of-freedom system is:
$$m\frac{d^2x}{dt^2} + c\frac{dx}{dt} + kx = F_0 cos\omega t, \qquad (2.1)$$
where $x(t)$ is the displacement from the equilibrium. Rewriting in terms of the damping ratio $\gamma = c/m$ and natural frequency $\omega_0 = \sqrt{k/m}$:
$$\frac{d^2x}{dt^2} + \gamma\frac{dx}{dt} + \omega_0^2 x = F_0 cos\omega t/m \qquad (2.2)$$

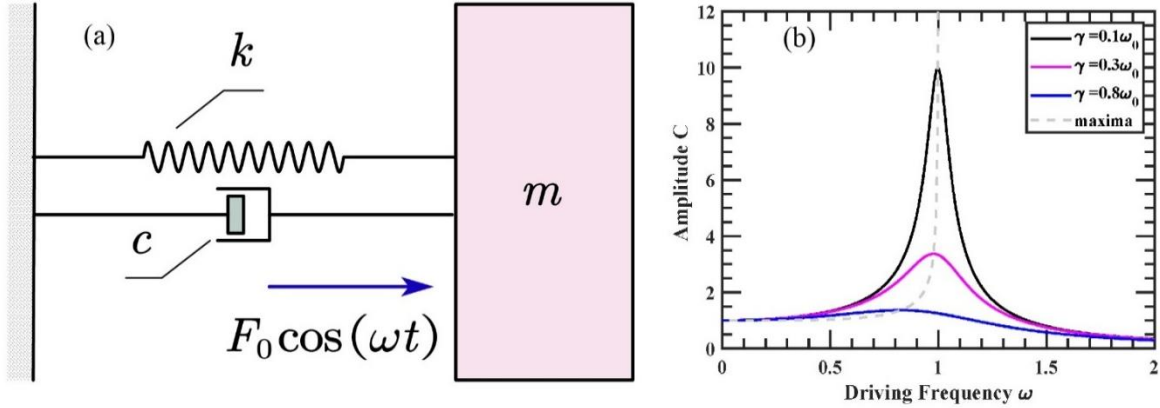

Figure 1. (a) Single-degree-of-freedom system with viscous damping subjected to forced vibration due to an external force acting on the mass. (b) Theoretical amplitude-frequency response of the system for $F_0 = 1\,N, m = 1\,kg$, and $\omega_0 = 1\,rad/s$. Resonance occurs at $\omega = \omega_{res}$, with peak amplitude determined by $\gamma$.

Under weak damping ($\gamma << \omega_0$) with $x(0) = x_0$ and $\dot{x}(0) = 0$, the general solution combines transient and steady-state components:

$$x(t) = \underbrace{x_0 e^{-\gamma t/2} cos(\omega_r t)}_{Transient} + \underbrace{C cos(\omega t + \theta_p)}_{Steady-State}, \qquad (3)$$

where $\omega_r = \sqrt{\omega_0^2 - \frac{\gamma^2}{4}}$ is the damped natural frequency and C and $\theta_p$ are the steady-state amplitude and phase shift (derived in Appendix A).

As $t \to \infty$, the transient term decays, leaving the steady-state solution. The steady-state solution of the system is given in Equation (3), with amplitude C and phase shift $\theta_p$ defined as:

$$C = \frac{F_0/m}{\sqrt{(\omega_0^2 - \omega^2)^2 + (\gamma\omega)^2}} \qquad (4)$$

$$\theta_p = arctan\frac{-\gamma\omega}{\omega_0^2 - \omega^2} \qquad (5)$$

Key characteristics of forced vibration: 1. Forced vibration is maintained by an external force rather than the system's intrinsic oscillations; 2. After the transient response dissipates, the system oscillates at the driving frequency rather than its natural frequency; 3. The phase difference between the driving force and the system's response depends on damping, driving frequency, and natural frequency; 4. Equation (4) predicts a maximum displacement at resonance:

$$C_{max} = \frac{F_0/m}{\gamma\sqrt{\omega_0^2 - \gamma^2/4}}.$$

At resonance, the driving frequency matches the system's resonance frequency: $\omega_{res} = \sqrt{\omega_0^2 - \gamma^2/2}$.

Figure 1(b) illustrates resonance, where the vibration amplitude peaks significantly when the driving frequency matches the resonance frequency (grey dashed line). This effect strongly depends on damping: lower damping results in a sharper resonance peak, as illustrated in the figure (solid line). This study emphasizes weakly damped systems due to their prevalence in natural and engineered systems (e.g., bridges and optical cavities). The simplified solutions (Equations 3–5) enable students to explore key concepts such as resonance experimentally, as detailed in Section 3.

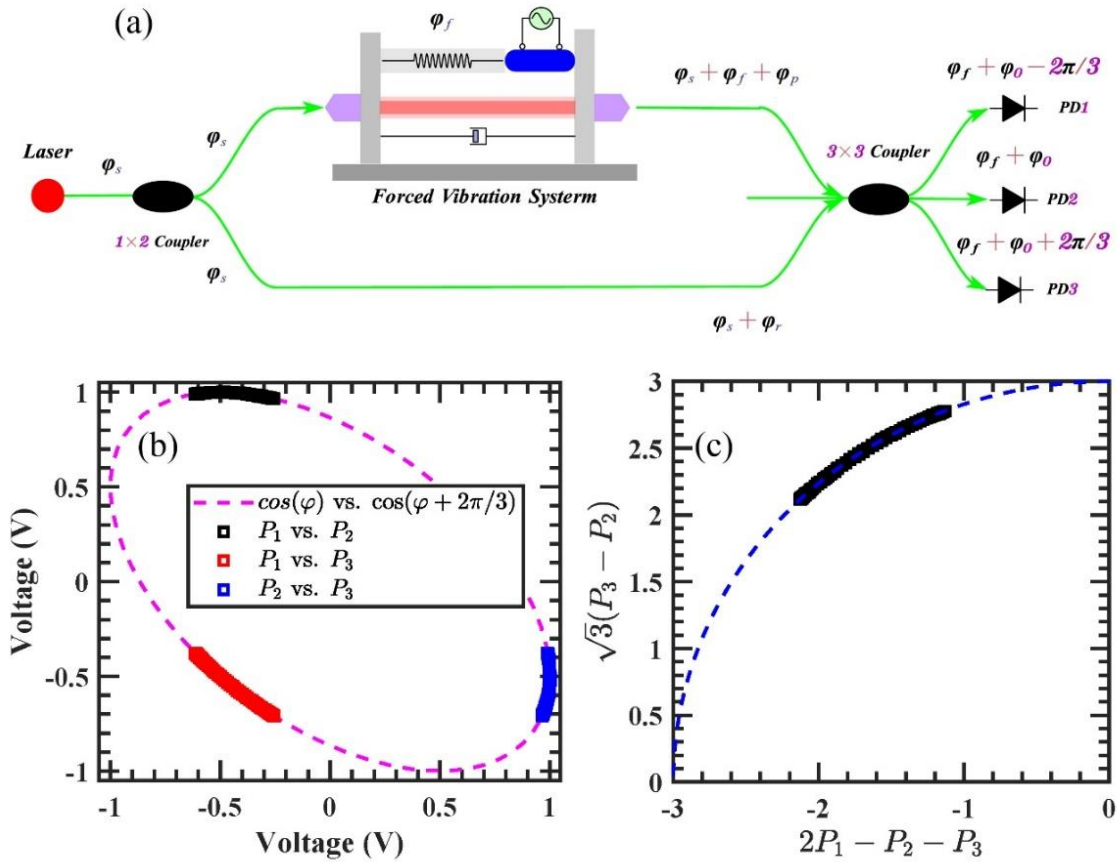

Figure 2. (a) Schematic diagram of the OFI. (b) Relationship between $P_1$, $P_2$, and $P_3$, where the signals form an ellipse. (c) Relationship between $\sqrt{3}\,(P_3 - P_2)$ and $2P_1 - P_2 - P_3$, where the signals form a circle. The phase $\varphi$ is determined using an inverse trigonometric function.

*2.2　Fiber Optic Laser Interferometer*

The OFI system (Figure 2(a)) integrates advanced optical fiber components to achieve high-precision displacement measurement. The key components include:

1. Laser source: A narrow-linewidth fiber laser (center wavelength: $\lambda = 1550\ nm$; linewidth <1 kHz) ensures frequency stability.
2. Beam splitter: A $1 \times 2$ fiber coupler divides the input beam into reference and probe beams.
3. Probe arm: The probe beam is collimated by a fiber collimator mounted on a forced vibration system. A kinematic mount ensures precise alignment of the collimated beam.
4. $3 \times 3$ fiber coupler: The $3 \times 3$ coupler recombines the reference and probe beams and produces three output signals with a fixed 120° phase difference, facilitating high-precision phase demodulation for directional displacement detection.
5. Detection module: Three photodetectors ($PD1$, $PD2$, $PD3$) convert optical signals into electrical outputs ($P_1, P_2, P_3$). A detailed description is provided in Appendix B.

The outputs of the PDs are expressed as:

$$P_1 = H + J\cos(\varphi - 2\pi/3), \tag{6}$$

$$P_2 = H + J\cos(\varphi), \tag{7}$$

$$P_3 = H + J\cos(\varphi + 2\pi/3) \tag{8}$$

where $H$ is the background intensity, $J$ is the signal amplitude, and $\varphi$ denotes the phase variation between the probe and reference paths.

Prior to the experiment, the phase relationship among $P_1$, $P_2$ and $P_3$ was verified, as shown in Figure 2(b), confirming the expected $120^o$ phase difference. Further analysis shows that $\sqrt{3}\ (P_3 - P_2)$ and $2P_1 - P_2 - P_3$ form an orthogonal signal pair (Figure 2(c)), enabling accurate phase calculation.

When the forced vibration system moves by a distance $x$, the optical path difference induces a phase shift given by:

$$\varphi = \varphi_f + \varphi_0$$

$$\varphi_f = 2\pi x/\lambda,$$

where $\varphi_f$ is the dynamic phase shift induced by displacement and $\varphi_0$ is the static initial phase difference between detection ($\varphi_P$) and reference ($\varphi_r$) paths, ideally remaining constant over time.

Using the $3 \times 3$ coupler outputs, the displacement can be accurately derived via the arctangent algorithm:

$$\varphi = arctan\ (\sqrt{3}\ (P_3 - P_2)/(2P_1 - P_2 - P_3\ )) \tag{9}$$

$$x = \varphi_f \lambda/2\pi. \tag{10}$$

This equation utilizes the phase differences among the three output signals to precisely calculate the displacement and determine its direction. Compared to conventional interferometry, this method not only measures the displacement direction but also offers higher temporal resolution. However, Equation (9) is valid only for single-frequency vibration signals. If multiple vibration components with comparable amplitudes exist, phase calculation errors may increase, leading to measurement inaccuracies.

## 3. Implementation

Before conducting the experiment, students received basic training in interferometric diagnostics, including optical alignment, data acquisition, and signal analysis. The experimental setup is shown in Figure 2(a). In this experiment, the spring constant was estimated to be approximately $k = (6.3 \pm 0.64) \times 10^6 \ N/m$, and the mass of the oscillating object was approximately $0.2 \pm 0.001 \ kg$, leading to a theoretical natural frequency of $\omega_0/2\pi \approx 896 \pm 45 \ Hz$. The system was driven by a piezoelectric transducer (PZT) [24], whose vibration frequency and amplitude were controlled by adjusting the input voltage and frequency. The PZT's maximum vibration frequency was 200 kHz, and at an input voltage of 10 V, the displacement amplitude was approximately 1 µm.

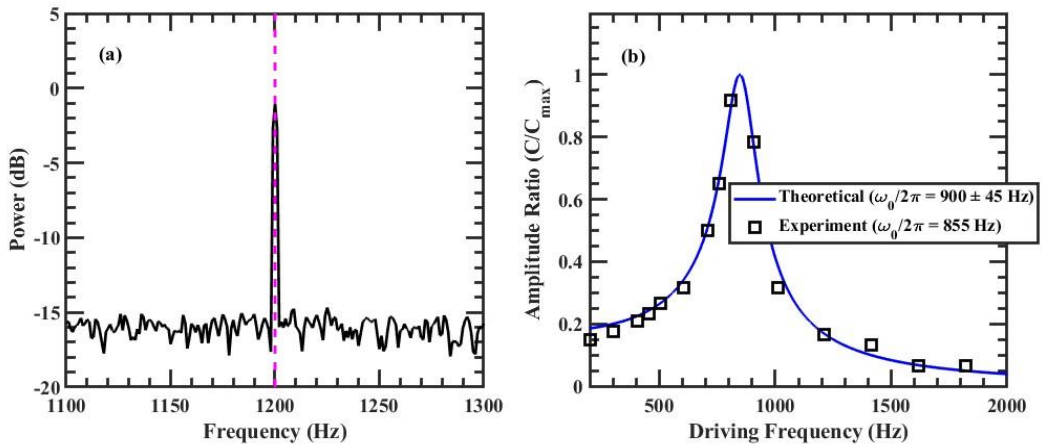

Figure 3. (a) power spectrum of the OFI-measured phase signal $\varphi_f$ at a driving frequency of 1200 Hz. (b) Normalized amplitude-frequency response obtained via OFI. Experimental data (black squares) are compared with theoretical predictions (blue line).

An OFI with a wavelength of 1550 nm was used to monitor the system's vibrations. The setup included two collimators: one fixed at the stationary end of the spring (Figure 2a) and the other attached to the oscillating object. By analyzing phase variations in the optical path, displacement measurements were obtained. To reduce the influence of transient responses, data analysis focused on measurements taken after $t = 10 \ s$.

Figure 3(a) presents the power spectrum of the OFI-measured phase signal $\varphi_f$ at a driving frequency of 1200 Hz (purple dashed line). A peak at 1200 Hz was observed, indicating agreement between the measured and driving frequencies and providing a basis for further analysis. To explore the system's resonance characteristics, the input frequency was scanned at a fixed 10 V voltage. Figure 3(b) shows the averaged results from 20 independent measurements, depicting the normalized displacement ratio $C/C_{max}$ as a function of driving frequency (black squares). Data fitting yielded a resonance frequency of $\omega_{res}/2\pi = 848\ Hz$, a natural frequency of $\omega_0/2\pi = 855$ Hz, and a damping coefficient of $\gamma = 158$ (blue solid line). The measured natural frequency (855 Hz) falls within the estimated theoretical range (900 ± 45 Hz), confirming consistency with the forced vibration model. Additionally, this method provides a reference for estimating the spring constant k.

## 4. Discussion

This study effectively demonstrated the feasibility of integrating OFI into forced vibration experiments. Traditional time-averaged interferometry faces challenges when dealing with directional ambiguity, whereas OFI's phase-sensitive detection (Figure 2c) provides unambiguous displacement tracking—a critical feature for teaching wave interference dynamics. The system's driving force, generated by a PZT, allowed precise control of frequency and amplitude through voltage modulation. The experimental results indicate the expected forced vibration behavior. By scanning the output frequency, we determined the natural frequency to be approximately 855 Hz, which is within the range of theoretical predictions. Despite damping and material uncertainties, the alignment between experimental and theoretical values underscores the robustness of OFI for educational applications. Future iterations could incorporate student-led uncertainty quantification exercises to foster analytical skills.

The student feedback and experimental data confirm the experiment's potential to enhance conceptual understanding and technical skills: 85% of students (n = 20) reported an improved grasp of phase-displacement relationships after analyzing quadrature signals ($\sqrt{3}\ (P_3 - P_2)$ vs. $2P_1 - P_2 - P_3$, Figure 2c), 40% expressed heightened interest in precision metrology, and 20% seeking additional lab time to refine measurements; A student team leveraged this work to develop an interferometric sensor optimization project funded by a university innovation grant. These findings align with broader calls to modernize classical mechanics labs through technology-enhanced inquiry [25].

## 5. Conclusion

This study demonstrates how OFI can bridge classical forced vibration theory and advanced optical metrology in educational settings. By enabling high-precision displacement measurements, the experiment reinforces core concepts such as resonance and damping while introducing students to real-world interferometric techniques. Preliminary observations indicate that structured hands-on practice—particularly in aligning interferometers and interpreting quadrature signals—deepens conceptual understanding and fosters problem-solving skills.

The modular "3+2" teaching design (3 hours of guided experiments + 2 hours of open exploration) accommodates diverse learners, allowing undergraduates to focus on foundational principles while graduate students tackle advanced signal processing tasks. Future studies could explore how early exposure to such interdisciplinary experiments impacts student retention in optics-related fields—a critical consideration for addressing workforce demands in photonics engineering. This approach aligns with global STEM education initiatives advocating for "technology-augmented labs" to enhance classical physics instruction.

**Acknowledgments**

This study was supported by the Natural Science Research Project of the Anhui Educational Committee (Grant no. 2024AH051100). The authors gratefully acknowledge the support of the Institute of Energy, Hefei Comprehensive National Science Center (Anhui Energy Laboratory) (No. 21KZS202).

## Appendix A: Derivation of Forced Vibration Solutions

I. Simple Harmonic Oscillator

When the damping coefficient $c = 0$ and the amplitude of the driving force $F_0 = 0$ in Equation (2), the system reduces to a simple harmonic oscillator. The governing differential equation becomes:

$$\frac{d^2 x}{dt^2} + \omega_0^2\, x = 0.$$

Here, $\omega_0$ is the natural frequency of the undamped system. We solve the equation in the complex domain:

$$\frac{d^2 \hat{x}}{dt^2} + \omega_0^2 \hat{x} = 0 \tag{I-1}$$

Here, $\hat{x}(t)$ represents the complex solution in the complex domain, and displacement of mass from equilibrium position $x(t)$ is its real part. Let us solve this step-by-step.

Trial Solution:

To solve Equation (I-1), assume a trial solution of the form: $\hat{x}(t) = A\, e^{\alpha t}$, where $A$ is a complex amplitude and α is a constant to be determined. Substituting this into (I-1), the equation becomes:

$$(\alpha^2 + \omega_0^2) A\, e^{i\alpha t} = 0.$$

Since $A\, e^{\alpha t} \neq 0$ for non-trivial solutions, we must have:

$$\alpha^2 + \omega_0^2 = 0$$

$$\alpha = \pm i\omega_0$$

Thus, the general solution is a linear combination of two independent solutions:

$$\hat{x}(t) = A_1\, e^{i\omega_0 t} + A_2\, e^{-i\omega_0 t},$$

where $A_1$ and $A_2$ are complex constants.

Ensuring a Real Solution:

Since the $x(t)$ must be real at all times, the coefficients must satisfy $A_2 = A_1^*$, where $A_1^*$ is the complex conjugate of $A_1$. Let $A_1 = C\, e^{i\theta_0}$, where C is a real amplitude and $\theta_0$ is a phase shift. Then:

$$\hat{x}(t) = C\, e^{i(\omega_0 t + \theta_0)} + C\, e^{-i(\omega_0 t + \theta_0)}. \tag{II-2}$$

Using Euler's formula, $e^{i\phi} + e^{-i\phi} = 2\cos(\phi)$, we get:

$$x(t) = 2C\cos(\omega_0 t + \theta_0\ ).$$

This is the real physical displacement.

Initial Conditions:

If $x(0) = x_0$ and $\dot{x}(0) = 0$,

$$x(0) = 2C\cos(\omega_0 \cdot 0 + \theta_0) = 2C\cos(\theta_0) = x_0$$

$$\dot{x}(0) = -2C\omega_0 \sin(\omega_0 \cdot 0 + \theta_0) = -2C\omega_0 \sin(\theta_0) = 0$$

Since $\omega_0 \neq 0$, we require:

$C = x_0/2, \theta_0 = 0$ or $C = -x_0/2, \theta_0 = \pi$.

The displacement of the simple harmonic oscillator is:

$$x(t) = x_0 \cos(\omega_0 t). \tag{I-3}$$

This solution describes oscillatory motion with amplitude $x_0$ and angular frequency $\omega_0$, satisfying the given initial conditions.

II. Underdamped Free Oscillation

We are tasked with understanding and verifying the solution to the underdamped free oscillation system described by the differential equation:

$$\frac{d^2 \hat{x}}{dt^2} + \gamma \frac{d\hat{x}}{dt} + \omega_0^2 \hat{x} = 0 \tag{II-1}$$

This equation arises when $F_0 = 0$ and the damping term is defined as $\gamma = c/m$, with the condition $0 < \gamma \ll \omega_0$. Here, c is the damping coefficient. Let us solve this step-by-step to confirm the solution provided.

Characteristic Equation:

To solve this second-order linear homogeneous differential equation, we assume a trial solution of the form: $\hat{x}(t) = A e^{\alpha t}$. Substituting this into (II-1), the equation becomes:

$$(\alpha^2 + \gamma\alpha + \omega_0^2)Ae^{i\alpha t} = 0 \Rightarrow \alpha = -\frac{\gamma}{2} \pm i\omega_r, \omega_r = \sqrt{\omega_0^2 - \frac{\gamma^2}{4}}.$$

The general solution is:

$$\hat{x}(t) = e^{-\frac{\gamma t}{2}}(Ae^{i\omega_r t} + A^* e^{-i\omega_r t}).$$

Referring to the simple harmonic oscillator solution method and taking the real part of $\hat{x}(t)$, we obtain:

$$x(t) = 2Ce^{-\frac{\gamma t}{2}} \cos(\omega_0 t + \theta_0).$$

The term $e^{-\frac{\gamma t}{2}}$ represents the exponential decay due to damping, while $\cos(\omega_r t)$ describes oscillations at the reduced frequency $\omega_r$.

Taking the real part and applying $x(0) = x_0, \dot{x}(0) = 0$:

$$C \approx x_0, \ \theta_0 \approx -\frac{\gamma}{2\omega_r} \approx 0.$$

Thus:

$$x(t) = x_0 e^{-\frac{\gamma t}{2}} \cos(\omega_r t). \tag{II-2}$$

This describes an oscillatory motion with frequency $\omega_r$, decreasing exponentially due to damping.

III. Underdamped Forced Oscillation

The full solution of Equation (2) includes:

1. Transient Solution ($x_c(t)$): The homogeneous solution (II-2), decaying as $t \to \infty$.

2. Steady-State Solution ($x_p(t)$): A particular solution driven by $F(t) = F_0 \cos(\omega t + \Delta)$.

Complex Forcing Term:

To solve the steady-state solution, it is typically easier to work with complex exponentials than trigonometric functions. We represent the real forcing term $F(t)$ as the real part of the complex forcing term: $\hat{F}(t) = F_0 e^{i(\omega t + \Delta)}$. For ease of calculation take $\Delta = 0$. The differential equation for the system, including damping and a restoring force, is given by:

$$\frac{d^2 \hat{x}_p}{dt^2} + \gamma \frac{d\hat{x}_p}{dt} + \omega_0^2 \hat{x}_p = \hat{F}(t)/m. \tag{III-1}$$

Trial Particular Solution:

The physical solution $x_p(t)$ will be the real part of $\hat{x}_p(t)$. Assume $\hat{x}_p(t) = Ae^{i\omega t}$, substituting $\hat{x}_p(t)$ into the equation:

$$(-\omega^2 + i\gamma\omega + \omega_0^2)Ae^{i\omega t} = \frac{F_0}{m} e^{i(\omega t)}.$$

Matching coefficients:

$$A = \frac{F_0}{m} \frac{1}{(\omega_0^2 - \omega^2) + i\gamma\omega}.$$

Amplitude and Phase:

Expressing $A = C e^{i\theta_p}$, the amplitude $C$ and phase shift $\theta_p$ of the response are:

$$C = \frac{F_0/m}{\sqrt{(\omega_0^2 - \omega^2)^2 + (\gamma\omega)^2}},$$

$$\tan \theta_p = \frac{-\gamma\omega}{\omega_0^2 - \omega^2}$$

Thus:

$$x_p(t) = \frac{F_0/m}{\sqrt{(\omega_0^2 - \omega^2)^2 + (\gamma\omega)^2}} \cos(\omega t + \theta_p) \tag{III-2}$$

Steady-State Behavior:

As $t \to \infty, x_c(t) \to 0$, leaving $x(t) \approx x_p(t)$.

## Appendix B: 3 ×x 3 Coupler

The schematic of the $3 \times 3$ fiber coupler is shown below.

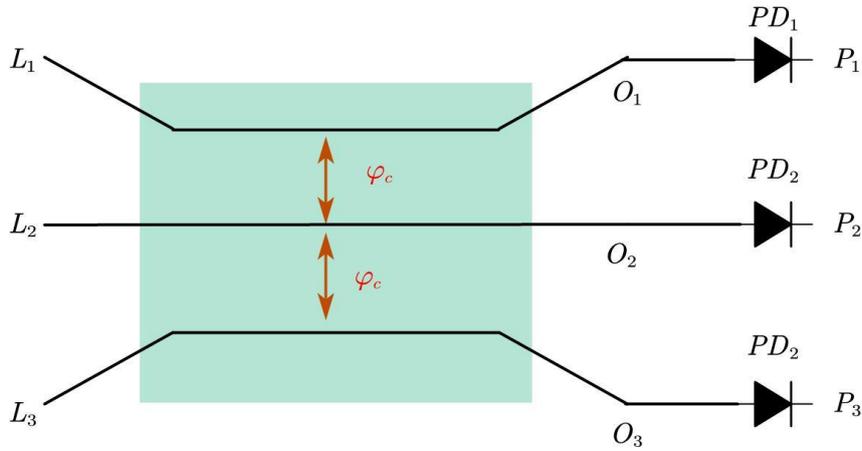

Figure 2. $3 \times 3$ Coupler Schematic.

The input signals are:

$L_3 = cos(\varphi_f + \varphi_s + \varphi_p)$, where $\varphi_f$ is the phase from forced vibration, $\varphi_s$ is the laser phase, and $\varphi_p$ is the probe path's initial phase.

$L_3 = cos(\varphi_s + \varphi_r)$, where $\varphi_r$ is the reference path's initial phase.

The phase shifts through the coupler are:

From $L_1$: $O_1\ (0^o), O_2\ (\varphi_c), O_3\ (2\varphi_c)$.

From $L_3$: $O_1\ (2\varphi_c), O_2\ (\varphi_c), O_3\ (0^o)$.

The resulting output signals are:

$$O_1 = E_1 \cos(\varphi_f + \varphi_s + \varphi_p) + D_1 \cos(\varphi_s + \varphi_r + 2\varphi_c),$$

$$O_2 = E_2 \cos(\varphi_f + \varphi_s + \varphi_p + \varphi_c) + D_2 \cos(\varphi_s + \varphi_r + \varphi_c),$$

$$O_3 = E_3 \cos(\varphi_f + \varphi_s + \varphi_p + 2\varphi_c) + D_3 \cos(\varphi_s + \varphi_r).$$

For an ideal $3 \times 3$ coupler, $\varphi_c = 120°$, and $E_1 = E_2 = E_3, D_1 = D_2 = D_3$.

After photodetection (PD), the outputs become:

$$P_1 = H + J \cos(\varphi - 2\pi/3),$$
$$P_2 = H + J \cos(\varphi),$$
$$P_3 = H + J \cos(\varphi + 2\pi/3),$$

where $H$ is background noise, $J$ is a constant, and $\varphi = \varphi_f + \varphi_0$, with $\varphi_0 = \varphi_p - \varphi_r$.

## References

[1] Babaei A. Forced vibrations of size-dependent rods subjected to: Impulse, step, and ramp excitations. Arch Appl Mech. 2021;91(5):2211-2223.
[2] Hubbard PG, Xu J, Zhang S, et al. Dynamic structural health monitoring of a model wind turbine tower using distributed acoustic sensing (DAS). J Civ Struct Health Monit. 2021;11(3):833-849.
[3] Lelas K, Pezer R. Modeling the amplitude and energy decay of a weakly damped harmonic oscillator using the energy dissipation rate and a simple trick[J]. European Journal of Physics, 2024, 46(1): 015004.
[4] Hong J, Gui L, Cao J. Analysis and experimental verification of the tangential force effect on electromagnetic vibration of PM motor. IEEE Trans Energy Conver. 2023;38(3):1893-1902.
[5] Shirai T, Kurosawa H. Clinical application of the forced oscillation technique. Intern Med. 2016;55(6):559-566.
[6] Brashier B, Salvi S. Measuring lung function using sound waves: Role of the forced oscillation technique and impulse oscillometry system. Breathe. 2015;11(1):57-65.
[7] Civalek Ö, Akbaş Ş D, Akgöz B, et al. Forced vibration analysis of composite beams reinforced by carbon nanotubes. Nanomater. 2021;11(3):571.
[8] Feynman R P, Leighton R B, Sands M, et al. The Feynman lectures on physics; vol. i[J]. Am J Phys. 1965, 33(9): 750-752.
[9] Liu Y, Baker F, He W, et al. Development, assessment, and evaluation of laboratory experimentation for a mechanical vibrations and controls course. Int J Mech Eng Educ. 2019;47(4):315-337.
[10] Pal S K, Panchadhyayee P. Determining viscosity of a liquid with smartphone sensors: A classroom-friendly approach using damped oscillations[J]. The Physics Teacher, 2025, 63(1): 64-65.
[11] Noh J J, Kim S, Kim J B. Measurement of Young's Modulus Using a Michelson Interferometer[J]. The Physics Teacher, 2023, 61(7): 618-620.
[12] Zhang G, Ge Q, Wang H, et al. Novel design of phase demodulation scheme for fiber optic interferometric sensors in the advanced undergraduate laboratory[J]. European Journal of Physics, 2022, 43(6): 065301.
[13] Feenstra L, Julia C, Logman P. A Lego® Mach–Zehnder interferometer with an Arduino detector[J]. Physics Education, 2021, 56(2): 023004.
[14] Santosa I E. A double beam Fabry–Perot interferometer for measuring laser wavelength[J]. European Journal of Physics, 2021, 42(3): 035301.
[15] Wong W O, Chan K T. Quantitative vibration amplitude measurement with time-averaged digital speckle pattern interferometry[J]. Optics & Laser Technology, 1998, 30(5): 317-324.
[16] Werth A, West C G, Sulaiman N, et al. Enhancing students' views of experimental physics through a course-based undergraduate research experience [J]. Phys Rev Phys Edu Res. 2023, 19(2): 020151.
[17] Dandridge A. Fiber optic sensors based on the Mach–Zehnder and Michelson interferometers. In: Udd E, Spillman Jr. WB, editors. Fiber optic sensors: An introduction for engineers and scientists. Wiley; 2024. p. 213-248.
[18] Zhang M, Gao J, Liu Z, et al. Fiber optic Michelson interference experimental course towards the cultivation of undergraduates majoring in optical engineering[J]. European Journal of Physics, 2023, 44(4): 045702.
[19] Lan T, Bai Z, Zhang H, et al. Measurement of electron and neutral particle densities using a two-color optical fiber interferometer. Revf Sc Instrum. 2024；95(10).
[20] Tien DM, Van NTH, Tounsi A, et al. Buckling and forced oscillation of organic nanoplates taking the structural drag coefficient into account. Comput Concrete. 2023;32(6):553-565.

[21] Lan T, Zhang S, Ding W, et al. Long-time measurements of line-integrated plasma electron density using a two-color homodyne optical fiber interferometer. Rev Sc Instrum. 2021；92(9).

[22] Karlo Lelas, Nikola Poljak, Dario Jukić; Damped harmonic oscillator revisited: The fastest route to equilibrium. Am. J. Phys. 1 October 2023; 91 (10): 767–775.

[23] Kontomaris S V, Malamou A. An elementary proof of the amplitude's exponential decrease in damped oscillations[J]. Physics Education, 2024, 59(2): 025030.

[24] Habib M, Lantgios I, Hornbostel K. A review of ceramic, polymer and composite piezoelectric materials. J Phys D Appl Phys. 2022;55(42):423002.

[25] Ivanović M, Milićević A K, Aleksić V, et al. Experiences and perspectives of Technology-enhanced learning and teaching in higher education–Serbian case [J]. Procedia Comput Sci. 2018, 126: 1351-1359.